\documentclass[sigconf, nonacm]{acmart}
\PassOptionsToPackage{table,dvipsnames,svgnames,x11names}{xcolor}

\usepackage{todonotes}
\usepackage{graphicx}   
\usepackage{tikz}
\usepackage{amsmath}
\usepackage{graphicx}
\usetikzlibrary{positioning, fit, calc, arrows,  shapes.multipart, arrows.meta}
\usepackage{enumitem}
\usepackage{algpseudocode}
\usepackage{algorithm}
\usepackage{forest}
\usepackage{placeins} 
\usepackage{float} 
\usepackage{graphicx} 
\usepackage{colortbl} 
\usepackage{pifont}   

\usepackage{graphicx}
\usepackage{caption}
\usepackage{subcaption}

\usepackage{listings}
\usepackage{pgfplots}
\usepackage[table]{xcolor} 

\newcommand\vldbyear{2025}
\newcommand\vldbworkshop{Applied AI for Database Systems and Applications (AIDB 2025)}
\newcommand\vldbauthors{\authors}
\newcommand\vldbtitle{\shorttitle} 
\newcommand\vldbavailabilityurl{}
\newcommand\vldbpagestyle{plain} 

\begin{document}
\title{Research Challenges in Relational Database Management Systems for LLM Queries}

\author{Kerem Akillioglu}
\affiliation{%
  \institution{University of Waterloo}
  \city{Waterloo}
  \country{Canada}}
\email{k2akilli@uwaterloo.ca}

\author{Anurag Chakraborty}
\affiliation{%
  \institution{University of Waterloo}
  \city{Waterloo}
  \country{Canada}}
\email{a8chakra@uwaterloo.ca}

\author{Sairaj Voruganti}
\affiliation{%
  \institution{University of Waterloo}
  \city{Waterloo}
  \country{Canada}}
\email{sairajv@uwaterloo.ca}

\author{M. Tamer {\"O}zsu}
\affiliation{%
  \institution{University of Waterloo}
  \city{Waterloo}
  \country{Canada}}
\email{tamer.ozsu@uwaterloo.ca}







\begin{abstract}
Large language models (LLMs) have become essential for applications such as text summarization, sentiment analysis, and automated question-answering. Recently, LLMs have also been integrated into relational database management systems to enhance querying and support advanced data processing. Companies such as Amazon, Databricks, Google, and Snowflake offer LLM invocation directly within SQL, denoted as \emph{LLM queries}, to boost data insights. However, open-source solutions currently have limited functionality and poor performance. In this work, we present an early exploration of two open-source systems and one enterprise platform, using five representative queries to expose functional, performance, and scalability limits in today’s SQL-invoked LLM integrations. We identify three main issues: enforcing structured outputs, optimizing resource utilization, and improving query planning. We implemented initial solutions and observed improvements in accommodating LLM powered SQL queries. These early gains demonstrate that tighter integration of LLM+DBMS is the key to scalable and efficient processing of LLM queries.
\end{abstract}

\maketitle

\pagestyle{\vldbpagestyle}
\begingroup\small\noindent\raggedright\textbf{VLDB Workshop Reference Format:}\\
\vldbauthors. \vldbtitle. VLDB \vldbyear\ Workshop: \vldbworkshop.\\ 
\endgroup
\begingroup
\renewcommand\thefootnote{}\footnote{\noindent
This work is licensed under the Creative Commons BY-NC-ND 4.0 International License. Visit \url{https://creativecommons.org/licenses/by-nc-nd/4.0/} to view a copy of this license. For any use beyond those covered by this license, obtain permission by emailing \href{mailto:info@vldb.org}{info@vldb.org}. Copyright is held by the owner/author(s). Publication rights licensed to the VLDB Endowment. \\
\raggedright Proceedings of the VLDB Endowment. 
ISSN 2150-8097. \\
}\addtocounter{footnote}{-1}\endgroup

\ifdefempty{\vldbavailabilityurl}{}{
\vspace{.3cm}
\begingroup\small\noindent\raggedright\textbf{VLDB Workshop Artifact Availability:}\\
The source code, data, and/or other artifacts have been made available at \url{\vldbavailabilityurl}.
\endgroup
}



\section{Introduction}
\label{sec:intro}
Recent advances in artificial intelligence (AI) have enabled large language models (LLMs) to offer enhanced processing capabilities for structured (relational) data \cite{llm-query}, as well as workloads that require access to both structured and unstructured data \cite{patel2024lotus, eleet, suql}. LLMs excel at row-level inference because they can reason over context and semantics instead of relying on exact string matches. This semantic capability has proven effective for some of the most important data management tasks, such as entity and schema matching \cite{entity, schema}, and data cleaning \cite{retclean} by overcoming the limitations of earlier approaches that completely relied on exact string matching.

 The promise of row-level LLM inference has led to its integration into database management systems (DBMS). Database vendors such as Amazon \cite{aws}, Databricks \cite{databricks}, MotherDuck \cite{atwal2024motherduck}, Google \cite{google}, and Snowflake \cite{snowflake} have accommodated the invocation of LLM inference within SQL queries. This integration extends relational DBMS capabilities by enabling advanced sentiment analysis and natural language processing, thus enriching query results with nuanced insights. We refer to such SQL queries that invoke LLMs as \emph{relational LLM queries}, or simply \emph{LLM queries}. 

 Despite their growing adoption by major enterprises, the integration of LLMs with DBMSs remains underexplored, and the incorporation of LLMs introduces additional challenges. For reference, enterprise LLM inference solutions expose data to third-party providers, and it incurs significant privacy risks for sensitive information. Organizations seeking to leverage LLM capabilities without compromising data confidentiality must serve models locally. 
 This paper demonstrates the functional challenges of current LLM+DBMS integrations and shows that, even when they function, the eminent systems are slow because they lack effective optimizations and they scale poorly. As an example, running the simple LLM query in Figure \ref{fig:llmproj} on only 17,000 rows takes 5 hours to complete using local inference on a single A100 GPU combined with an open source DBMS; and it would take around 12 days to process one million rows in our single GPU setup. Hence, closing the functionality and efficiency gaps is critical for DBMSs to scale AI-enhanced workloads.


In this work, we provide an early exploration of processing LLM queries on existing systems through analyzing two open-source systems and one enterprise system from functionality and performance aspects. We use a fixed set of LLM queries in our analysis (see Section \ref{sec:queries}). We demonstrate that existing systems have difficulty executing many of the queries in this set, and we analyze the underlying challenges and potential solutions. Our contributions are the following:
\begin{enumerate}
    \item We present an evaluation of emerging systems that integrate LLM capabilities into SQL querying. Our evaluation uses open source solutions PostgreSQL with pgAI \cite{pgAI} and DuckDB with FlockMTL \cite{flockmtlpaper}, and an enterprise solution in MotherDuck \cite{atwal2024motherduck}. For each case, we examine functionality and overall performance, and analyze why queries are not able to run or have poor performance. 

    \item Our study provides practical insights and recommendations on what opportunities exist for query optimization for queries that involve LLM invocations, and what inference-based optimizations can be implemented to meet the specific requirements of analytical queries. 

\end{enumerate}

\section{Relational LLM Queries}
\label{sec:queries}
In our study, we focus on the queries outlined by Liu et al. \cite{llm-query} since they are a good representation for embedding LLM invocations at various points within SQL queries. Each query is identified by a unique name and characterized by its distinct LLM invocation strategy. Below, we summarize the five representative queries we use:

\textbf{Q1: LLM Projection} – This query invokes the LLM in the \verb+SELECT+ clause to project a transformed output from input text fields. The purpose is to derive insights directly from the provided textual data.
\begin{figure}[H]
    \centering
    \begin{lstlisting}
SELECT LLM("Recommend movies for the user based on {movie information} and {user review}", 
     m.movie_info, r.review_content)
FROM reviews r
JOIN movies m ON r.rotten_tomatoes_link ==
m.rotten_tomatoes_link
    \end{lstlisting}
    \caption{\label{fig:llmproj}  LLM Projection Query.}
\end{figure}

\textbf{Q2: LLM Filter} – Here, the LLM is invoked in the \verb+WHERE+ clause to evaluate and filter rows based on semantic criteria. The LLM function determines if the input text meets a specified condition, thereby controlling which records are included in the result set.
        \begin{figure}[H]
            \centering
            \begin{lstlisting}
SELECT m.movie_title
FROM Movies m
JOIN Reviews r ON r.rotten_tomatoes_link =
m.rotten_tomatoes_link
WHERE LLM("Analyze whether this movie would be suitable for kids based on {movie information} and {user review}", m.movie_info, r.review_content) == "Yes"
AND r.review_type == "Fresh"
            \end{lstlisting}
            \caption{\label{fig:llmfilter}  LLM Filter Query.}

        \end{figure}
  
\textbf{Q3: Multi-LLM Invocation} – This query combines two LLM invocations: one  in \verb+SELECT+ clause to generate primary outputs and the other in the \verb+WHERE+ clause to filter the results based on additional content suitability criteria. The purpose is to refine the final output by sequentially applying multiple LLM functions.
  \begin{figure}[H]
    \centering
    \begin{lstlisting}
SELECT LLM("Recommend movies for the user based on {movie information} and {user review}", m.movie_info, r.review_content) AS recommendations
FROM Movies m
JOIN Reviews r ON r.rotten_tomatoes_link =
m.rotten_tomatoes_link
WHERE LLM("Analyze whether this movie would be suitable for kids based on {movie information} and {user review}", m.movie_info, r.review_content) == "Yes" 
AND r.review_type == "Fresh"
    \end{lstlisting}
    \caption{\label{fig:multillm}  Multi-LLM Query.}

\end{figure}

\textbf{Q4: LLM Aggregation} – The LLM is called to assign satisfaction ratings for reviews for each movie title to qualitatively measure average sentiment for overall customer feedback. Then these numerical scores are aggregated using an average function. This approach synthesizes qualitative data into a quantitative summary metric.
  \begin{figure}[H]
    \centering
    \begin{lstlisting}
SELECT AVG(LLM("Rate a satisfaction score between 0 (bad) and 5 (good) based on {review} and {info}: ",r.review_content, m.movie_info)) as AverageScore
FROM reviews r
JOIN movies m ON r.rotten_tomatoes_link =
m.rotten_tomatoes_link
GROUP BY m.movie_title

    \end{lstlisting}
    \caption{\label{fig:llmaggr}  LLM Aggregation Query.}

\end{figure}
  
\textbf{Q5: RAG} – This query implements a Retrieval-Augmented Generation (RAG) approach by first retrieving relevant context via a similarity search and then using the LLM to generate an answer. The LLM invocation here enhances response generation by leveraging externally retrieved contextual data.
\begin{figure}[H]
    \centering
    \begin{lstlisting}
SELECT LLM("Given the following {context}, answer this question", 
     VectorDB.similarity_search(s.question), 
     s.question)
FROM squad s
WHERE s.is_impossible == False;
    \end{lstlisting}
    \caption{\label{fig:llmrag}  RAG Query.}

\end{figure}

\section{Systems Testing}
\label{sec:testing}
Our objectives are two-fold: (1) we want to understand what is required to be able to execute the queries discussed in the previous section, and (2) we want to test the performance of executing them and understand the factors that affect query performance. As we discuss below, most of these queries cannot be executed using these open-source systems. Some of the issues can be addressed by careful engineering designs, but others require new approaches that require research.




\subsection{Testing Setup}

Our setup uses a local inference model: Meta’s LLaMA 3.1 with 8B parameters \cite{llama}. This model is served with both Ollama and vLLM to understand the role of the model serving engine's impact on functionality and performance of LLM+DBMS integrations. As noted above, the open source systems we use are  PostgreSQL 16 with pgAI 0.8.0 and DuckDB v1.1.4 with FlockMTL v0.2.0 \cite{FlockMTL}. We run our experiments on a Linux server equipped with a single NVIDIA A100 80GB PCIe GPU, an Intel Xeon® Platinum 8380 CPU with 160 cores. We also test an enterprise system, MotherDuck, through using its API and MotherDuck Prompt() \cite{promptmotherduck} functionality which uses OpenAI GPT4o-mini \cite{openai2024gpt4technicalreport} as the LLM endpoint, and we use their built-in embedding \cite{emebeddingmotherduck} function.

Our queries (Section \ref{sec:queries}) and the datasets are the same as those in Liu et al. \cite{llm-query}: Rotten Tomatoes \cite{rotten} and Stanford Question Answering Dataset (SQuAD) \cite{squad}.  We had to limit the size of our datasets as explained in Section \ref{sec:intro}, because applying LLMs to large datasets incurs significant computational costs. We used the size of the \textit{movies} table of the Rotten Tomatoes dataset as our limit row number for the \textit{ reviews} and SQuAD data, which is 17,712 rows, and we randomly sample from these datasets. Rotten Tomatoes is used for queries 1-4 as it has a relational structure, and SQuAD is used for Q5 since question \& answering datasets are natively compatible with RAG because they are context-rich and semantically searchable. For RAG queries on open source systems, we utilize the Stella embedding model \cite{stella}.







\subsection{Functional Testing}
Table \ref{tab:rotten_functional} shows which systems are capable of running which queries, and we discuss the related research challenges in more detail in Section \ref{sec:challenges}.
Query 1 runs successfully on all tested systems since it simply summarizes previously retrieved rows and does not incur additional challenges. In contrast, queries 2, 3, and 4 fail in both FlockMTL\footnote{After sharing results with the FlockMTL team, they have enabled structured output when using OpenAI-compatible and Ollama providers.} and pgAI due to structured output problems that are discussed further in Section \ref{sec:constrained}. Specifically, queries 2 and 3 require the LLM to strictly output "Yes", and any deviation from this format prevents proper filtering. Similarly, query 4 demands the LLM to provide only a numeric output between 1 and 4 to compute an average. However, the LLM consistently adds extra text, which causes execution failures. MotherDuck, utilizes GPT-4o-mini for inference and successfully executes queries 1–4 due to OpenAI's structured, constrained decoding capabilities \cite{openaiStructuredOutputs}. To leverage vLLM's structured decoding capabilities through XGrammar \cite{xgrammar}, we modified FlockMTL to work with vLLM \cite{bentoml2025structured}, and ran queries on FlockMTL. This modification enabled the successful execution of queries 2 and 3, but not query 4 (LLM aggregation). The problem with query 4 stems from a structured output and query planner mismatch. Even with constrained decoding, FlockMTL's \texttt{llm\_reduce}, FlockMTL's LLM-aggregation function, returns an untyped text blob rather than the structured numeric type the planner needs for \verb+AVG()+, so the operator and optimizer cannot align. Query 5 also suffers from query planning issues and fails to run both on MotherDuck as well as FlockMTL. We traced this issue in DuckDB's query planner where the HNSW index lookup fails to trigger and instead performs a cross join between the two vector embedding tables being referenced. This leads to the query running out of memory due to the large intermediate table being materialized. pgAI is the only system that could execute query 5 because we manually enforced the optimal query plan (see Figure \ref{fig:query-plans}) to make it run, and the pgvector extension is tightly integrated into the pgAI environment. First, we applied the filter \texttt{s.is\_impossible == False} to narrow down the data, and then performed the similarity search (filter pushdown \& top-k pushdown). This returns only the top 3 relevant contexts for each question before making the LLM call. This fix is specific to this query because we had to alter its plan to make it run. Broader planning issues and requirements are discussed in Section \ref{sec:queryplans}.

\begin{table}
\centering
\renewcommand{\arraystretch}{1.2} 
\setlength{\tabcolsep}{9pt} 
\begin{tabular}{|l|c|c|c|c|c|}
\hline
\rowcolor{gray!20} 
\textbf{Systems} & \textbf{Q1} & \textbf{Q2} & \textbf{Q3} & \textbf{Q4} & \textbf{Q5} \\
\hline
\textbf{pgai-ollama} & \ding{51} & \ding{55} & \ding{55} & \ding{55}  & \ding{51} \\
\hline
\textbf{flockmtl-ollama} & \ding{51} & \ding{55} & \ding{55} & \ding{55} & \ding{55}  \\
\hline
\textbf{flockmtl-vllm} & \ding{51} & \ding{51} & \ding{51} & \ding{55} & \ding{55}  \\
\hline
\textbf{motherduck-gpt} & \ding{51} & \ding{51} & \ding{51} & \ding{51} & \ding{55}  \\
\hline
\end{tabular}
\vspace{0.1cm}
\caption{\textbf{Query success/failure for each system running Queries 1-5 from Section \ref{sec:queries}. \ding{51} denotes success, \ding{55} denotes failure. }}
\label{tab:rotten_functional}
\end{table}

\subsection{Performance Testing}
\label{sec:perf}
\begin{figure}
        \centering
        \includegraphics[width=\linewidth]{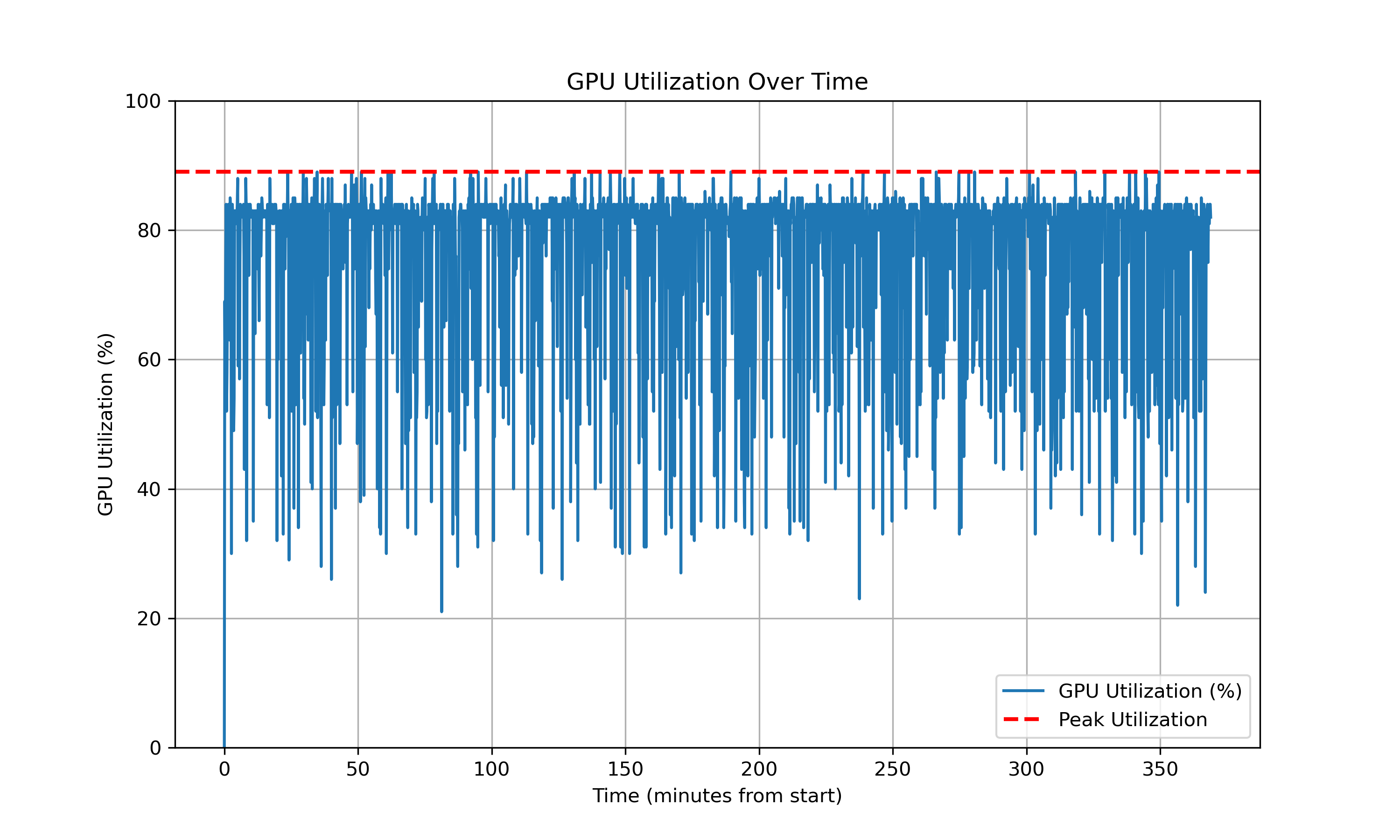}
        \caption{FlockMTL (Ollama) GPU Utilization for Q1}
        \label{fig:flock_Q1_ollama}
    \end{figure}

\begin{figure}
        \centering
        \includegraphics[width=\linewidth]{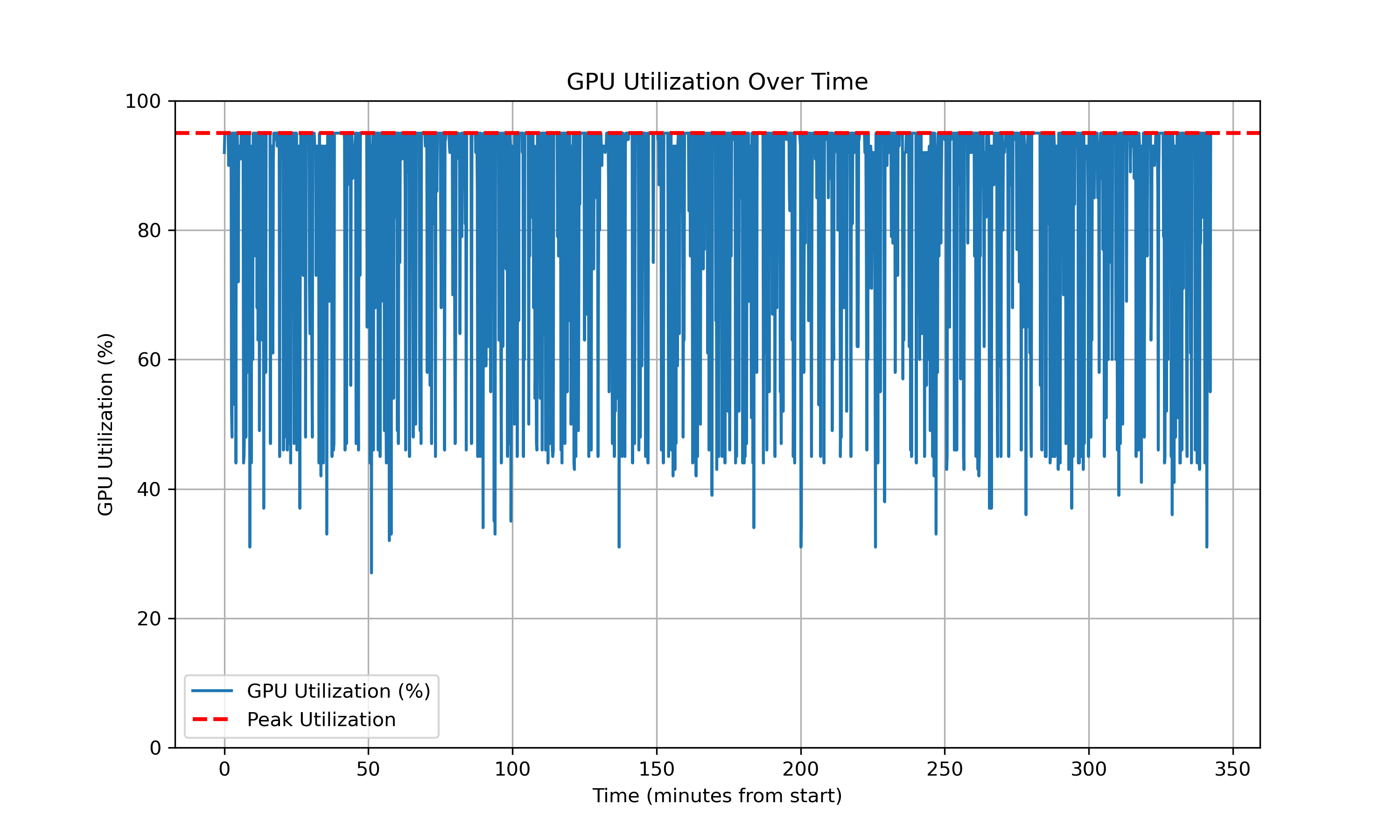}
        \caption{FlockMTL (vLLM) GPU Utilization for Q1}
        \label{fig:flock_Q1}
    \end{figure}

\begin{figure}
        \centering
        \includegraphics[width=\linewidth]{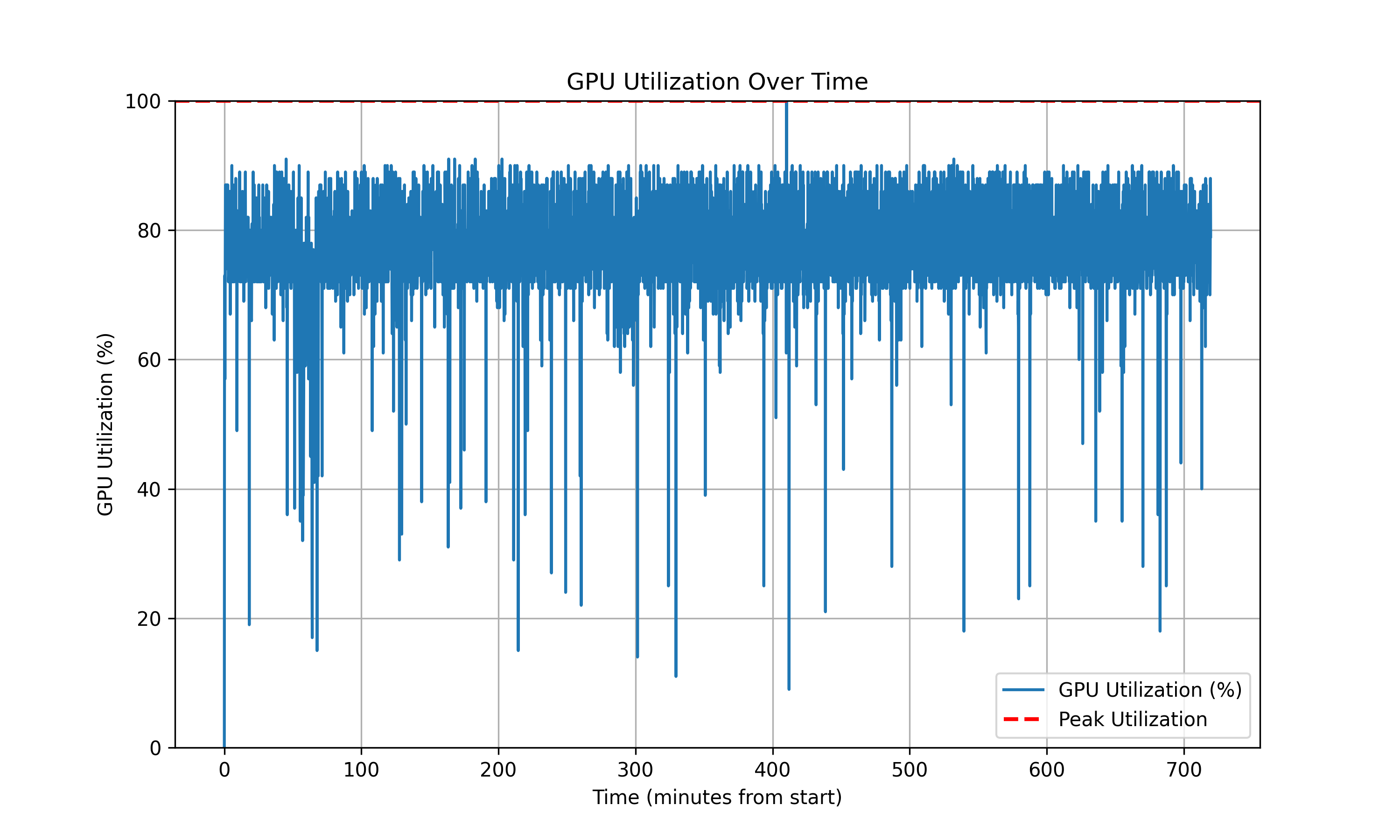}
        \caption{pgAI (Ollama) GPU Utilization for Q1}
        \label{fig:pgai_Q1}
    \end{figure}

\begin{figure}
        \centering
        \includegraphics[width=\linewidth]{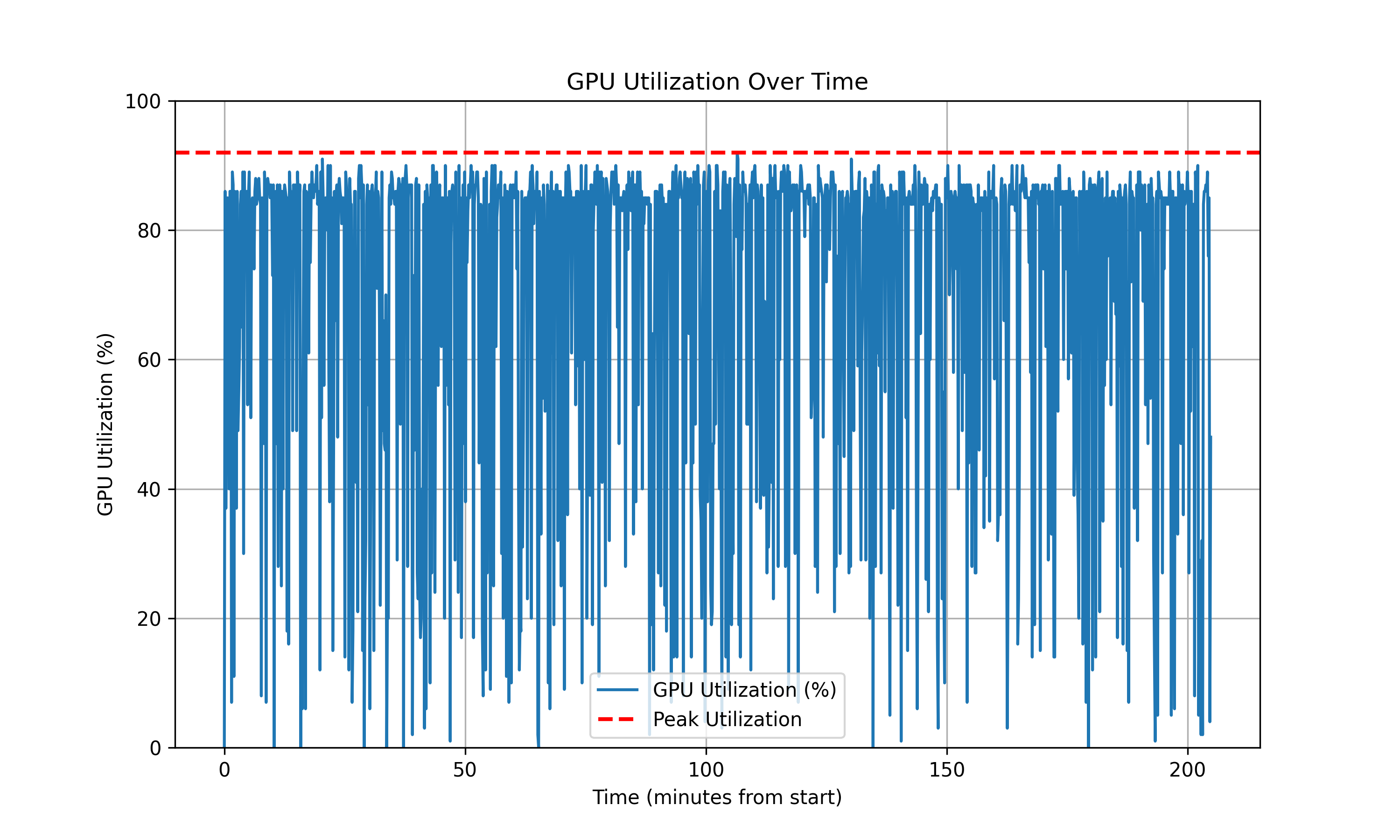}
        \caption{pgAI (Ollama) GPU Utilization for Q5}
        \label{fig:pgai_Q5}
    \end{figure}

In this section, we report query latencies and resource utilization for the studied workload. Table~\ref{tab:rotten_random} lists the latencies for each query, and Figures ~\ref{fig:flock_Q1_ollama}-\ref{fig:pgai_Q5} show GPU utilization graphs for a selected set of queries. As FlockMTL and pgAI follow different execution strategies, we first analyze how these strategies influence overall performance. We then present MotherDuck’s results separately since it is a closed enterprise service running on proprietary hardware, which is not directly comparable to the open-source systems.


\begin{table}[ht]
\centering
\renewcommand{\arraystretch}{1.2} 
\setlength{\tabcolsep}{7 pt} 
\begin{tabular}{|l|c|c|c|c|c|}
\hline
\rowcolor{gray!20} 
\textbf{Systems} & \textbf{Q1} & \textbf{Q2} & \textbf{Q3} & \textbf{Q4} & \textbf{Q5} \\
\hline
\textbf{pgai-ollama} & 719.5 & - & - & - & 204.9 \\
\hline
\textbf{flockmtl-ollama} & 370.0 & - & - & -& -  \\
\hline
\textbf{flockmtl-vllm} & 342.4 & 447.6 & 617.2 & -& -  \\
\hline
\end{tabular}
\vspace{0.1cm}
\caption{\textbf{Query Latency (minutes) for each system. pgai-ollama denotes the relational extension (pgai) and inference engine (ollama) being used. - denotes the query was not able to run. }}
\label{tab:rotten_random}
\end{table}

The general execution strategy of FlockMTL for all its LLM functions is to batch multiple rows in a single prompt request by taking the total context window of the model (e.g., 128,000 tokens for Llama models), subtracting the tokens needed for the system prompt, and then sending enough rows until the window limit is reached. It then instructs the model to output exactly 1 row for each input row in JSON format. In practice, this approach does not work well since the LLM is not robust enough to follow these instructions for hundreds of rows. Another issue is that the LLM may output multiple lines of rows, and sometimes not even a single row for an input row. These issues make it infeasible to track which output row from the LLM output maps to an input row. To fix these issues we modified FlockMTL's code to keep the batch size for each prompt to exactly one row, which ensured the queries run to completion. 
However, in terms of GPU utilization, this approach is inefficient since it only sends 1 input row as a prompt request at-a-time which is why FlockMTL's runtime numbers 
are an order of magnitude slower (with both Ollama and vLLM) compared to the enterprise system MotherDuck. From Figure \ref{fig:flock_Q1_ollama}, we observe that the peak GPU utilization across the query 
runtime for Q1 with Ollama (85\%) is less than the peak GPU utilization with vLLM (Figure \ref{fig:flock_Q1}, 95\%), which reflects in the runtime numbers (Table 
\ref{tab:rotten_random}).

On the other hand, in PostgreSQL, parallel query execution enhances performance by distributing the work among multiple worker processes. As the data volume increases, PostgreSQL automatically assigns more workers to process the data concurrently, and pgAI is designed similarly. However, when we attempted to run Q1 as a single block, we observed that pgAI's implementation and its corresponding UDF (User-Defined Function) try to initiate subtransactions within these parallel workers. Since PostgreSQL does not allow subtransactions during parallel operations, an error\footnote{We opened an issue on the pgAI GitHub repository; however, as of this writing, it has remained unanswered for more than a week.} gets triggered, and the query does not run. 


To resolve this issue, we forced the query to run in a single-process context where subtransactions are permitted. While this fix resolves the error, processing one row at-a-time results in poor performance and resource utilization. To improve efficiency, we implemented an optimization to batch requests in pgAI and observed significant performance gains. Without this optimization, processing a single row for query 1 took around 4 seconds on average; and after applying it, per row processing time dropped to 2.5 seconds. Our implementation also showed its effect on GPU utilization. When there's no batch optimization of requests, GPU utilization fluctuates dramatically, as can be seen in Figure \ref{fig:pgai_Q5}. With optimization, we recorded a stable and higher GPU utilization graph compared to processing other queries, which shows an average utilization of 76\% and a median of 75\%.



The enterprise solution (motherduck-gpt) performs best across all queries, as shown in Table \ref{tab:motherduck_tab}. The reason is largely because it leverages the OpenAI API for inference rather than relying on local inference. This approach takes advantage of the state-of-the-art performance and efficiency of enterprise systems and hardware as opposed to having to optimize local inference on a single GPU. In addition, MotherDuck further improves speed by concurrently sending up to 256 requests to the model provider \cite{blogmotherduck}. By parallelizing these requests, the system significantly reduces overall latency and accelerates processing, leading to consistently lower query response times.

\begin{table}[ht]
\centering
\renewcommand{\arraystretch}{1.2}
\setlength{\tabcolsep}{7 pt}
\begin{tabular}{|l|c|c|c|c|c|}
\hline
\rowcolor{gray!20}
\textbf{System} & \textbf{Q1} & \textbf{Q2} & \textbf{Q3} & \textbf{Q4} & \textbf{Q5} \\
\hline
\textbf{motherduck-gpt} & 16.2 & 4.3 & 13.4 & 8.3 & -- \\
\hline
\end{tabular}
\vspace{0.1cm}
\caption{\textbf{Query latency (minutes) for the proprietary MotherDuck platform.}}
\label{tab:motherduck_tab}
\end{table}

\section{Research Challenges}

\label{sec:challenges}
Existing systems are still in the development stage, and many research challenges remain to be addressed. These challenges span multiple aspects, including structured output handling, resource utilization, and query planning.

\subsection{Structured Outputs}
\label{sec:constrained}
Due to their non-deterministic nature, LLMs can generate syntactically inconsistent outputs, and for LLM queries, this can lead to the generation of schema-breaking text that disrupts downstream processing. For example, in an ``LLM-Aggregation'' query, the model must return a single number between 1 and 5; if LLM instead outputs anything else, the \verb+AVG+ SQL function cannot parse the value, and the query fails to execute. For any system that supports relational LLM queries, structured outputs are not a technical enhancement, it’s a fundamental requirement to run queries. To address this issue, there are three possible solutions, each with its own trade-offs: fine-tuning the model, prompt engineering and constrained decoding.

Fine-tuning is the process of supplementing a pretrained model with domain-specific training on small datasets to perform better for particular tasks. Once tuned, the fine-tuned model captures nuanced domain semantics with minimal runtime overhead. It has also been effective on tabular data for applications like data synthesis and privacy protection \cite{tabby, tabdatasynth}. The trade-offs are as follows: collecting schema-accurate examples can be costly, parameter-efficient tuning consumes significant time and compute resources \cite{qlora}, in the case of a schema change, another tuning pass is required, and closed-source models cannot be tuned at all.

Prompt engineering seeks optimal task performance by tailoring prompts to a specific model–dataset pair \cite{prompteng}. It can be complemented with system prompts that add global instructions \cite{sysprompts}. This technique adds virtually no runtime cost and is model-agnostic because behavior changes are just text edits. Its success, however, is probabilistic and context-specific. Despite having semantically equivalent prompts for the same question, performance can vary significantly \cite{promptvar}. Also, each task, schema, or domain requires its own tailored prompt, whose effectiveness can decline in long or multi-turn contexts \cite{lost,moretoken}.

Constrained decoding is the technique to modify LLM’s token-generation process so that each subsequent token is restricted to choices that preserve the required output structure. Doing so, the generated text can be ensured to adhere to constraints like high-level templates through using regular expressions \cite{beurer2023}, context-free grammars \cite{willard2023,xgrammar}, or a combination of both \cite{guidance_library}. It has proven to be effective in text-to-SQL translation tasks \cite{picard}, and has been used in practical applications like vLLM \cite{xgrammar, outlines_library}. Advantages of using structured decoding removes the need for fine-tuning, any additional post-processing, ad-hoc parsing, retrying and prompting on top of LLMs \cite{guidingllmsrightway}. Although effective in ensuring correctness, it increases the computational overhead and incurs challenges with batch and parallel processing as discussed in Section \ref{sec:util}.

For our work, our initial solution for this problem was to use prompt engineering. Once the dataset grew, our optimized prompts yielded inconsistent outputs that prevented efficient parsing.  We ruled out fine-tuning due to its high cost of curating large volumes of schema-perfect examples and spending hours of GPU compute. Instead, we used constrained decoding: giving the model an explicit grammar that systematically forces every answer to match the table schema and requires no retraining.

\subsection{Resource Utilization} 
\label{sec:util}
During inference, batching strategies play a key role in resource utilization as well as query latency. Frequent data transfers between the CPU and GPU reduce overall GPU utilization, and the absence of an effective batching strategy largely explains the spikes observed in Figures \ref{fig:flock_Q1_ollama}, \ref{fig:flock_Q1}, and \ref{fig:pgai_Q5}. Our workload requires a combination of GPU-CPU operations, and the inference engine uses the GPU to process the input prompt (prefill) and generate output tokens (decode). A good scheduling design for relational LLM queries should aim for maximum GPU utilization, and involve sending multiple concurrent asynchronous requests to the inference backend instead of blocking API calls. Asynchronous parallel requests to the inference engine ensure that the GPU bound output token generation operation overlaps with the CPU bound operations. This allows the GPU to remain active by generating tokens for a concurrent request while waiting for output validation of a prior request. 

Also, when batching requests, the inference engine should separate batches that require constrained generation from those that do not. Due to structured decoding, the logit masks for each decoding step must be generated and validated on the CPU to constrain the output. This results in each decoded token being passed to the CPU one at-a-time. When output token validation occurs on the CPU, the GPU remains idle in the blocking API request case. Therefore, mixing constrained and non-constrained requests in the same batch leads to performance degradation as the overhead of constrained requests slows down the non-constrained ones \cite{anyscale}.


On top of batching requests, row level batching is a potential strategy as FlockMTL's design suggests. However, it comes with practical challenges, as mentioned in Section \ref{sec:perf}. When using this approach, the number of input tokens and the output size must be carefully calculated to fit within the model’s context window. Ideally, the best performance would come from batching multiple rows into a single request while ensuring each row's output is handled separately and running multiple requests asynchronously. This would depend on new methods being developed beyond constrained decoding that can send multiple rows of input to an LLM, and return separate, per-row output back to the database.

\newcommand{\PlanQueryOne}{%
\begin{tikzpicture}[
    node distance = 0.5cm and 0.5cm,
    every node/.style = {draw, rounded corners, align=center, font=\sffamily},
    >=Stealth,               
    scale=0.9, transform shape 
]
\node (proj)  {Projection \\ (llm\_complete)};
\node (hash)  [below = of proj] {Hash Join};
\node (scanR) [below = of hash] {Scan\\Reviews};
\node (scanM) [right = of scanR] {Scan\\Movies};
\draw[->] (hash)  -- (proj);
\draw[->] (scanR) -- (hash);
\draw[->] (scanM.north) |- (hash.east);
\end{tikzpicture}}

\newcommand{\PlanQueryTwo}{%
\begin{tikzpicture}[
    node distance = 0.5cm and 0.5cm,
    every node/.style = {draw, rounded corners, align=center, font=\sffamily},
    >=Stealth,
    scale=0.9, transform shape
]
\node (proj)  {Projection};
\node (filter) [below = of proj] {Filter \\ (llm\_filter)};
\node (hash)   [below = of filter] {Hash Join};
\node (scanR)  [below = of hash] {Scan\\Reviews};
\node (scanM)  [right = of scanR] {Scan\\Movies};
\draw[->] (filter) -- (proj);
\draw[->] (hash)   -- (filter);
\draw[->] (scanR)  -- (hash);
\draw[->] (scanM.north) |- (hash.east);
\end{tikzpicture}}

\newcommand{\PlanQueryThree}{%
\begin{tikzpicture}[
    node distance = 0.5cm and 0.5cm,
    every node/.style = {draw, rounded corners, align=center, font=\sffamily},
    >=Stealth,
    scale=0.9, transform shape
]
\node (proj)  {Projection \\ (llm\_complete)};
\node (filter) [below = of proj] {Filter \\ (llm\_filter)};
\node (hash)   [below = of filter] {Hash Join};
\node (scanR)  [below = of hash] {Scan\\Reviews};
\node (scanM)  [right = of scanR] {Scan\\Movies};
\draw[->] (filter) -- (proj);
\draw[->] (hash)   -- (filter);
\draw[->] (scanR)  -- (hash);
\draw[->] (scanM.north) |- (hash.east);
\end{tikzpicture}}

\newcommand{\PlanQueryFour}{%
\begin{tikzpicture}[
    node distance = 0.5cm and 0.5cm,
    every node/.style = {draw, rounded corners, align=center, font=\sffamily},
    >=Stealth,
    scale=0.9, transform shape
]
\node (aggr)  {Aggregate};
\node (proj)  [below = of aggr] {llm\_reduce (per group)};
\node (filter) [below = of proj] {Hash Group By };
\node (hash)   [below = of filter] {Hash Join};
\node (scanR)  [below = of hash] {Scan\\Reviews};
\node (scanM)  [right = of scanR] {Scan\\Movies};
\draw[->] (proj) -- (aggr);
\draw[->] (filter) -- (proj);
\draw[->] (hash)   -- (filter);
\draw[->] (scanR)  -- (hash);
\draw[->] (scanM.north) |- (hash.east);
\end{tikzpicture}}

\newcommand{\PlanQueryFive}{%
\begin{tikzpicture}[
    node distance = 0.5cm and 0.5cm,
    every node/.style = {draw, rounded corners, align=center, font=\sffamily},
    >=Stealth,
    scale=0.9, transform shape
]
\node (proj)  [below = of aggr] {Projection \\ (llm\_complete)};
\node (filter) [below = of proj] {Similarity Search };
\node (hash)   [below = of filter] {Filter};
\node (scanR)  [below = of hash] {Scan\\SQuAD};
\draw[->] (filter) -- (proj);
\draw[->] (hash)   -- (filter);
\draw[->] (scanR)  -- (hash);
\end{tikzpicture}}

\begin{figure*}[t]
  \centering
  \resizebox{1.0\textwidth}{!}{
    \begin{minipage}{0.192\linewidth}\centering
      \PlanQueryOne
      \caption*{Query 1}
    \end{minipage}\hfill
    \begin{minipage}{0.192\linewidth}\centering
      \PlanQueryTwo
      \caption*{Query 2}
    \end{minipage}\hfill
    \begin{minipage}{0.192\linewidth}\centering
      \PlanQueryThree
      \caption*{Query 3}
    \end{minipage}\hfill
    \begin{minipage}{0.1922\linewidth}\centering
      \PlanQueryFour
      \caption*{Query 4}
    \end{minipage}\hfill
    \begin{minipage}{0.192\linewidth}\centering
      \PlanQueryFive
      \caption*{Query 5}
    \end{minipage}
  }
  \caption{Query plans for relational-LLM queries}
  \label{fig:query-plans}
\end{figure*}

\subsection{Query Plans}
\label{sec:queryplans}

There are two main strategies in integrating LLM functions in the query planning stage. Databricks \cite{databricks}, Amazon Redshift \cite{aws}, and pgAI use UDFs to call LLMs, while FlockMTL uses custom C++ functions that make LLM calls.  

Having UDFs allows LLM queries to run directly through SQL statements without the need to integrate external tools into existing workflows. However, these UDFs are treated as "blackboxes" by the query planner and no optimization rules are applied when making the LLM inference call. Figure \ref{fig:query-plans} shows the query plans for the relational LLM queries, and they are similar across all the systems we tested. While certain optimizations such as filter pushdowns are observed for queries 2 and 4 (Figure \ref{fig:query-plans}), optimizations within the LLM function call such as reordering of the column attributes being referenced based on cardinality estimates are not considered. 

Current relational DBMS optimizers leave a lot of performance on the table by ignoring the cost of LLM invocations. The planner can recover that performance by factoring in the LLM prompts embedded in UDFs and other custom functions. To minimize latency and cost, the optimizer should consider CPU-GPU resource allocation and execution ordering, batching, KV-cache reuse, and prompt overlap as first-class objectives. Guided by these insights, such as prefix-cache sharing across low cardinality columns \cite{llm-query}, the planner can reorder input rows so tuples with similar attribute values appear together. Thus, prompt prefixes can be shared, overlapping text can be batched, attention states can be reused, and pre-fill overhead can be reduced. Incorporating the true cost of LLM calls can close a gap in current relational optimizers and unlock substantial untapped performance. 

\section{Related Work}
\label{sec:related}
As LLMs are increasingly being applied in different areas, a growing research community is addressing challenges to apply LLMs to structured tabular data tasks \cite{llmtabular}. This community tackles problems such as generating new table columns or features \cite{nam2024optimizedfeaturegenerationtabular}, data cleaning \cite{retclean}, table understanding through representation learning \cite{tablerep}, and semantically enriching table content \cite{tabenrich}. Even though this community's work is highly relevant, we are not concerned with such tasks, instead, we aim to explore how LLM-enhanced workloads can be accommodated within DBMSs as we conduct an early exploration for the open-source systems that enable the execution of SQL queries that invoke LLMs.

To efficiently process AI-powered analytical queries, LLMs are used for unstructured (text-centric) \cite{docetl, dai2024uqe, liu2024declarative} and structured (relational) \cite{llm-query} workloads, both separately and in combination  \cite{patel2024lotus, eleet, suql}. Researchers introduce optimizations like prompt prefix-sharing and row-reordering \cite{llm-query}, designing declarative querying primitives \cite{patel2024lotus, eleet} and physical operators \cite{papotti} to process data efficiently with LLMs. What we do in our work complements these systems to efficiently support their workloads as it is orthogonal to these solutions, and can be adopted with such frameworks.

RAG enhances retrieval by fetching only the most relevant data from large datasets, keeping LLM prompts within length limits while grounding the model’s output in essential information \cite{ji2024target, LMRAG}. We are not concerned with using LLMs for RAG operations. Instead, we aim to optimize the DBMS+LLM interaction by enabling LLM-powered SQL queries to be performed efficiently.

\section{Conclusion}
\label{sec:future}
Running LLM queries is attractive because they extend traditional SQL and can greatly enrich analytics in relational DBMSs. Yet executing these queries efficiently and practically is difficult, especially when the model must run locally to avoid exposing sensitive data and maintain privacy. Our early exploration reveals key functionality and performance challenges, as well as the trade-offs of executing LLM queries inside relational DBMSs.

Key challenges are enforcing structured outputs, optimizing batching, and improving query planning. We highlight trade-offs among structured output techniques, show how insufficient batching decreases the GPU utilization, and demonstrate that planner-execution mismatches can harm functionality and performance. We also highlight the need for incorporating LLM costs into query planning to maximize resource utilization. Our initial solutions demonstrate measurable improvements in LLM+DBMS integration. We believe that LLM queries might benefit from adopting different approaches, such as approximate query processing techniques, and may require rethinking relational DBMS query execution strategies to maximize GPU and model efficiency.

\bibliographystyle{ACM-Reference-Format}


\end{document}